\documentclass[aps,prl,twocolumn,twoside,floatfix,a4paper,showpacs]{revtex4}

\usepackage{graphicx}
\usepackage{textcomp}
\usepackage{amssymb}\usepackage{amsmath}\usepackage{amsthm}
\DeclareSymbolFont{AMSb}{U}{msb}{m}{n}
\DeclareMathSymbol{\N}{\mathbin}{AMSb}{"4E}
\DeclareMathSymbol{\Z}{\mathbin}{AMSb}{"5A}
\DeclareMathSymbol{\R}{\mathbin}{AMSb}{"52}
\DeclareMathSymbol{\Q}{\mathbin}{AMSb}{"51}
\DeclareMathSymbol{\I}{\mathbin}{AMSb}{"49}
\DeclareMathSymbol{\C}{\mathbin}{AMSb}{"43}

\newcommand{\la}{\langle}
\newcommand{\ra}{\rangle}
\newcommand{\be}{\begin{equation}}
\newcommand{\ee}{\end{equation}}
\newcommand{\bea}{\begin{eqnarray}}
\newcommand{\eea}{\end{eqnarray}}
\newcommand{\pd}{\partial}

\newcommand{\lt}{\left(t\right)}

\begin{document}
\title{Decoherence in a nonequilibrium environment}
\author{Joseph Beer}
\author{Eric Lutz}

\affiliation{Department of Physics, University of Augsburg, D-86135 Augsburg, Germany}
\date{\today}

\begin{abstract}
We consider a quantum harmonic oscillator coupled to a general nonequilibrium environment. We show that the  decoherence factor can be expressed in terms of a  measurable  effective temperature, defined via a generalized fluctuation-dissipation relation.  We further propose  a simple experimental  scheme to determine the time-dependent effective temperature in a linear Paul trap with engineered reservoirs. Our  formalism allows quantitative description of nonequilibrium decoherence in the presence of an arbitrary number of  non-Markovian noise sources in a unified manner.
\end{abstract}
\pacs{03.65.Yz}
\maketitle

Recent theoretical and experimental work on  decoherence has focused on the paradigmatic model of a single quantum harmonic oscillator (system) linearly coupled to an ensemble of harmonic oscillators (equilibrium reservoir) \cite{giu96,sch07}. The interaction with the  reservoir leads to the dynamical suppression  of interference fringes which arise in coherent superpositions of oscillator states. Decoherence, therefore, plays a fundamental role in the description of the quantum-to-classical transition \cite{zur91,zur03}. Consider for example a typical cat state, $\psi_{\mbox{\scriptsize cat}}(x) = [\psi_-(x) +\psi_+(x)]/\sqrt{2}$, consisting of a linear superposition of two Gaussian wave packets $\psi_\pm(x)$ of width $\sigma$ and separated by a distance $d  \,(\gg\sigma)$. Then, for temperatures much larger than the oscillator frequency, the interference-fringe contrast, or decoherence factor,  decays exponentially with time according to \cite{zur91,zur03},
\be
\label{1}
{\cal D}(t) = \exp( -D \,d^2 t)\ .
\ee
The coherence time $t_c= 1/(D\, d^2)$ is here inversely proportional to the amplitude $D$ of the thermal fluctuations and to the square of the spatial separation  of the wave packets. The Gaussian dependence of the decoherence factor  on the distance $d$  is responsible for the  nonobservation of quantum superposition of macroscropic states. For an equilibrium reservoir,  the  coefficient $D$ is related to the friction constant $\eta$ via the Einstein relation, $D = \eta T$, where $T$ is the temperature (we use natural units for which $\hbar=k_B=1$). For a given coupling strength $\eta$, the decoherence rate is hence directly controlled by the temperature of the reservoir. The form  of the equilibrium decoherence factor \eqref{1} has been successfully verified in experiments involving mesoscopic superpositions of a radiation field inside a microwave cavity \cite{bru96} and of motional states of a single trapped ion \cite{mya00}. In most cases of practical interest, however, a quantum system is subjected to many noise sources with different amplitudes and correlation times, corresponding de facto to a nonequilibrium environment.

In this paper we investigate nonequilibrium decoherence in the presence of many noise sources  that act independently on the quantum system. The concrete system we have in mind is an ultracold ion confined in a large scale rf trap and subjected to different fluctuating electric  fields \cite{kie02,des06}, but our approach is more general. We model each noise source by an equilibrium  harmonic reservoir with given coupling constant, temperature and correlation time, and take non-Markovian effects fully into account. Despite the complexity of the problem, we show that the nonequilibrium decoherence factor can be expressed in terms of a simple measurable quantity. To accomplish this task, we use recent results from nonequilibrium statistical physics and map the coupling to $M$ different equilibrium reservoirs to the coupling to a single nonequilibrium reservoir that we characterize with the help of a time-dependent effective temperature. We define the latter via  a  nonequilibrium extension of the fluctuation-dissipation relation and  suggest an experimental  method to determine  it  in a linear Paul trap.  The notion of effective temperature has been originally introduced in the description of glassy dynamics \cite{cug93,cug97}; it satisfies the expected properties of a temperature, such as a zeroth law (two interacting observables that evolve on the same time-scale have the same effective temperature) and has been measured in many nonequilibrium systems, for instance  in classical spin \cite{her02} and colloidal \cite{abo04} glasses, as well as in granular systems \cite{aba08} and ageing polymer glasses \cite{ouk09}. We  apply our   results to the description of decoherence of a   trapped ion interacting with two  different amplitude reservoirs, treating  the case of real and artificially engineered reservoirs \cite{poy96}. We show   that in  both  instances nonequilibrium decoherence can be slowed down in a controlled manner as compared to the corresponding equilibrium situation, and  explain how this behavior can be  understood and predicted with the help of  the effective temperature.

\begin{figure}
 \centering
     \includegraphics[width=.47\textwidth]{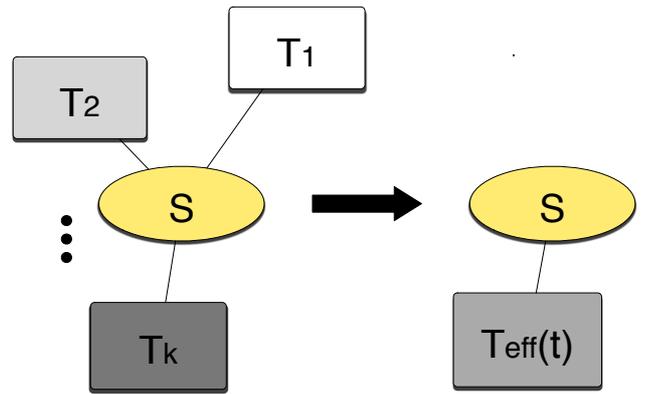}
  \caption{Schematic representation of the open quantum system. A complex environment, made of  different equilibrium reservoirs, is mapped to  a single nonequilibrium reservoir, described by an effective time-dependent temperature.}
\label{fig1}
 \end{figure}

The Hamiltonian of  an oscillator  $S$ linearly coupled to $M$ arbitrary harmonic reservoirs $R_k$ is of the form 
\begin{equation}
H =  H_{S} + \sum_{k=1}^M \sum_{j=1}^{N_k}  \frac{p_{kj}^2}{2m_{kj}}+\frac{m_{kj}\omega_{kj}^2}{2} \left( q_{kj}- \frac{c_{kj} x}{m_{kj} \omega_{kj}^2}\right)^2,
\label{P2}
\end{equation}
where $H_S= p^2/(2m) + m\omega^2 x^2/2$ is the Hamiltonian of the system oscillator with the usual notation  and $c_{kj}$ denote coupling constants. Each equilibrium  reservoir $R_k$  is fully specified by a temperature $T_k$ and a spectral density function $J_k(\omega) = (\pi/2) \sum_j c_{kj}^2/(m_{kj}\omega_{kj}) \, \delta (\omega -\omega_{kj})$ that characterizes the coupling to the system; we stress that we do not restrict ourselves to the Ohmic regime, $J_k(\omega)\sim \omega$, but include frequency-dependent damping.  We describe the time evolution of the  harmonic system with the exact quantum  Langevin equation,
\be
m \ddot x(t) + \int_0^t dt' \eta(t-t') \dot x(t') + m\omega^2 x(t) = F(t)-\eta(t) x(0) \ ,
\ee
which follows from the total Hamiltonian \eqref{P2} \cite{wei93}. Here the friction kernel $\eta(t) = \sum_k \eta_k(t)$ and the fluctuation force $F(t) = \sum_k F_k(t)$ are  given by the sum of the respective contributions  of  the individual reservoirs.  Under very general conditions (essentially that the reservoir is only weakly perturbed by the motion of  the system \cite{cal83}), any  equilibrium reservoir can be modeled by an ensemble of harmonic oscillators \cite{rei01}.
Since all reservoirs $R_k$ are in thermal equilibrium, the correlation functions $C_k(t)=\la F_k(t)F_k(0) + F_k(0)F_k(t)\ra/2$ of the thermal forces and the friction kernels  $\eta_k(t)= 2\int_0^\infty d\omega J_k(\omega) /(\pi m \omega) \cos \omega t $  are connected via a fluctuation-dissipation relation.  In the following, we consider   the physically relevant regime for decoherence studies, in which  temperature is larger than the oscillator frequency \cite{bru96,mya00}.  The fluctuation-dissipation relation is then given in Laplace space by $C_k[s] = T_k \eta_k[s]$ \cite{cal51}, indicating that, at equilibrium, energy damping  and thermal fluctuations evolve on the same time scale; we note that
this equation can be interpreted as defining the equilibrium temperature $T_k$ of each reservoir. The above   relation   only holds at thermal equilibrium. However, away from equilibrium, we can extend the   fluctuation-dissipation relation and define a frequency-dependent effective temperature  as  \cite{zam05},
\be
\label{7}
T[s] = \frac{C[s]}{\eta[s]}  = \frac{\sum_k T_k\eta_k[s]}{\sum_k \eta_k[s]}\ ,
\ee
where we have introduced the correlation function $C[s]=\sum_k C_k[s]$ of the total fluctuating force $F[s]$.
We mention that a low-temperature extension of the effective temperature has been proposed in Ref.~\cite{cug98}. With the help of  Eq.~\eqref{7}, we can now map the complex problem of a quantum system coupled to $M$ arbitrary equilibrium reservoirs to that of a  system interacting with a  single nonequilibrium reservoir characterized by a time-dependent effective temperature $T_\text{eff}(t) = \int_{0^-}^t dt' T\left(t'\right)$ (see Fig.~\ref{fig1}). In the case of a single equilibrium reservoir, $M=1$, we obtain $T(t) = T_1 \delta (t)$ and the effective temperature reduces to the reservoir temperature, $T_\text{eff}= T_1$, as expected. Conversely, we note that a general nonequilibrium environment, with arbitrary $\eta(t)$ and $C(t)$, can be approximated with $M$ equilibrated reservoirs \cite{zam05}. The effective temperature is a useful quantity to characterize the violation of the fluctuation-dissipation relation in nonequilibrium systems. We will show below that it also provides deep insight into nonequilibrium decoherence.

We  turn to the dynamics of decoherence of a cat state in a general nonequilibrium environment. We focus on the regime of microscopic separation that corresponds to  long coherence times, and investigate the influence of a nonequilibrium reservoir which evolves on similar or shorter time scales. 
The  importance of nonequilibrium effects on short time scales  can be made clear by considering the coherent transport of ions in a segmented linear Paul trap: fast, nonadiabatic  transport of the order of a few oscillation periods (i.e. a fraction of the coherence time) has been experimentally shown to drive  ions  into nonequilibrium excited states \cite{hub08}. The accurate description of short-time nonequilibrium effects appears therefore crucial. Decoherence processes are conveniently analyzed  in Wigner phase space \cite{zur91}.
The  Wigner function of  a Gaussian   cat state is given by the sum of two classical contributions $W_\pm(x,p)$ and a quantum interference term $W_\text{int}(x,p)$,
 $W\left(x,p\right)=W_-\left(x,p\right)+W_+\left(x,p\right)+W_\text{int}\left(x,p\right)$.
During  decoherence, the interference term progressively decays  as a function of time. We quantify the disappearance of the contrast of the interference fringes by means of  the peak-to-peak ratio  between the interference and the classical terms of the Wigner function  \cite{paz93},
\begin{equation}
\exp{\left(-A_\text{int}\right)}=\frac{W_\text{int}\left(x,p\right)\vert_\text{peak}}{2\sqrt{W_-\left(x,p\right)\vert_\text{peak}W_+\left(x,p\right)\vert_\text{peak}}} \ .
\label{8}
\end{equation}
The nonequilibrium peak-to-peak ratio can be directly expressed  as a function of the effective temperature  \eqref{7} in the following way.
We first note that for  high temperatures, the dynamics of the system  is dominated by  momentum diffusion driven by the fluctuating force $F(t)$. We can thus describe the evolution  of the Wigner function by the approximate Fokker-Planck equation, where free evolution and dissipation terms are neglected \cite{zur91},
\be
\label{9}
\frac{\pd W(x,p,t)}{\pd t}= \eta(t) x \frac{\pd W}{\pd p} + \sigma^2_{pp}(t) \frac{\pd^2 W}{\pd p^2} \ .
\ee
Here $\sigma^2_{pp}(t) = 2 \int_0^t dt'\int_0^{t'} dt'' C(t'-t'')$ is the momentum variance which describes the width of the Wigner function. The attenuation factor can  now be readily evaluated from the solution of Eq.~\eqref{9} and we find   $A_\text{int}(t)  = d^2 \sigma^2_{pp}(t) /2$. According to the definition \eqref{7} of the effective temperature, the noise correlator is further given by the convolution $C(t) = \int_0^t dt' \eta(t-t') T(t')$. We therefore  obtain,
\be
 A_\text{int}(t)= d^2 \int_0^t dt'\int_0^{t'} dt''\eta\left(t'-t''\right)T\left(t''\right)  (t-t')\ .
\label{10}
\ee
The above nonequilibrium peak-to-peak ratio $\exp{\left(-A_\text{int}\right)}$ is the direct extension of the familiar equilibrium decoherence factor \eqref{1} to which it reduces when $T(t) = T \delta(t)$; it completely characterizes decoherence  of a quantum system coupled to a general nonequilibrium environment, with arbitrary $\eta(t)$ and $C(t)$.  As in the equilibrium case, for a given coupling strength the decoherence rate is determined by the (effective) temperature of the nonequilibrium reservoir. 

Let us now illustrate our theoretical results by analyzing the concrete situation of a quantum oscillator that interacts  with a fast and a slow reservoir. This is the simplest non-trivial out-of-equilibrium environment \cite{zam05}.  We begin by discussing   'real' environments before treating the case of 'engineered' environments in ion traps in the next section.  We specify the fast (Markovian) reservoir by a temperature $T_f$ and a delta-correlated friction kernel, $\eta_f\left( t\right) =\eta_f \delta\left( t\right) $, and the slow (non-Markovian) reservoir by a temperature $T_s$ and an exponentially-correlated  kernel, $\eta_s\left( t\right) =(\eta_s/\tau) \exp{\left( -t/\tau\right)}$, with correlation time $\tau$. The effective temperature \eqref{7} is then given by
\begin{equation}
 T[s]=\frac{T_f\eta_f \left( s\tau+1\right) +T_s\eta_s}{\eta_f\left( s\tau+1\right) +\eta_s}\ ,
\label{11}
\end{equation}
and the corresponding effective temperature in the time-domain reads, with $\eta=\eta_f+\eta_s$,
\begin{equation}
 T_\text{eff}\lt=\frac{\eta_f T_f+\eta_s T_s}{\eta}+\frac{\eta_s}{\eta}\left(T_f-T_s\right)\exp{\left(-\frac{\eta}{\eta_f}\frac{t}{\tau}\right)}\ .
\label{12}
\end{equation}

\begin{figure}[t]
 \centering
  \includegraphics[width=0.45\textwidth]{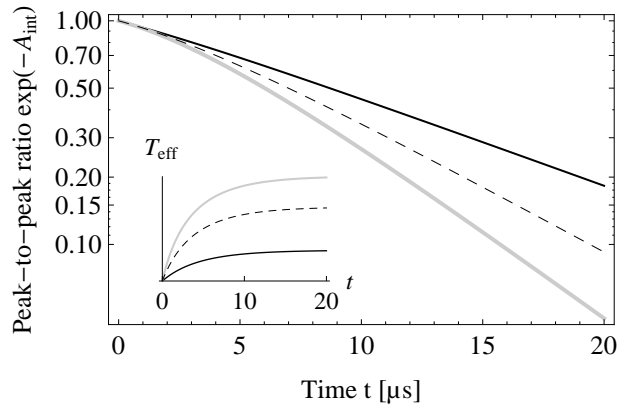}
\caption{Interference-fringe contrast for a nonequilibrium two-reservoir system simulated by two random electric fields $E_1(t)$ and $E_2(t)$.  Nonequilibrium decoherence is  slower (grey) or faster (black)  than equilibrium decoherence (dashed) when the temperature $T_f\sim \la E_1^2\ra $ of the fast  reservoir is higher  or lower than the temperature $T_s\sim \la E_2^2\ra $ of the slow reservoir. The dashed curve is the  reference line corresponding to  equal noise amplitudes $\la E_1^2\ra =\la E_2^2\ra $. The  time-dependent effective temperature (shown in the inset)  predicts this behavior: it is smaller than the reference temperature  in the first case, while it is larger in the second. The effective temperature can be determined from measurements of  the motional heating rate for trap frequencies $\omega/(2\pi)$ varying between 1MHz and 100MHz. Parameters are: $d=7$nm, $\sigma=0.12$nm, $\tau=5 \mu$s, $\la E_1^2\ra=10^{-9} \mbox{V}^2\mbox{m}^{-2}$ and $\la E_2^2\ra=6\cdot 10^{-10} \mbox{V}^2\mbox{m}^{-2}$. }
 \label{fig2}
 \end{figure}

\noindent The effective temperature \eqref{12} evolves monotonously from the fast reservoir temperature $T_f$ at $t=0$ to the weighted average $T_{\infty}=\left(\eta_f T_f+\eta_s T_s\right)/\eta$ of the two temperatures at large times, $t\gg\left(\eta_f/\eta\right)\tau$. The exact peak-to-peak ratio $\exp{\left(-A_\text{int}\right)}$  \eqref{8} for the nonequilibrium two-reservoir model is plotted in Fig.~\ref{fig2} for three differents sets of reservoir parameters. We observe that the pace of decoherence can be controlled by properly tuning the reservoir variables: nonequilibrium decoherence is  slower (faster) than equilibrium decoherence, which corresponds to  $T_f=T_s$, when the temperature of the slow reservoir is lower (higher) than the temperature of the fast reservoir. We can therefore dramatically affect the decoherence properties of the system  by, for instance, simply exchanging the correlation times of the two reservoirs, while keeping their respective temperatures constant. This is a nontrivial non-Markovian effect. The effective temperature $T_\text{eff}$  offers an intuitive explanation for this  behavior. Following  Eq.~\eqref{12}, we note that  when $T_f>T_s$ ($T_f<T_s$), the two-reservoir configuration  is equivalent to a nonequilibrium reservoir with a decreasing (increasing) effective temperature. The  approximate expression \eqref{10}  of the interference-fringe contrast then immediately shows that decoherence  is slowed down (or accelerated).

Linear Paul traps are  promising candidates for quantum computation and an invaluable tool for the investigation of the dynamics of decoherence \cite{lei03}. Due to their weak interaction with the ambient reservoir, and the resulting long coherence time, they offer the unique possibility of exploring various decoherence scenarios generated by the coupling to well-defined engineered reservoirs \cite{poy96}. The coupling to an equilibrium amplitude reservoir can thus be simulated with the help of a fluctuating electric field that acts on the charged ion \cite{mya00}. In this case, the temperature  of the reservoir is given by the variance of the field. 
The nonequilibrium  two-reservoir configuration can be effectively simulated  with two random electric fields $E_1(t)$ and $E_2(t)$ with corresponding correlation times. The correlation function of the superimposed noisy fields is given by $C^E(t) =e^2 \la E_1^2 \ra \delta(t)+(e^2/\tau) \la E_2^2\ra  \exp(-t/\tau)$ and is identical to the two-reservoir correlation function $C(t)$ with the identification $e^2 \la E_1^2 \ra = \eta_f T_f$ and $e^2 \la E_2^2 \ra =\eta_s T_s$. Using the definition \eqref{7} of the effective temperature, we then find $T^E[s]=T+C^E[s]/\eta^E[s]\simeq C^E[s]/\eta^E[s]$, since the influence of the ambient reservoir can be neglected during the duration of the experiment. We note that the damping kernel $\eta^E[s] = \eta$ is here constant, in contrast to the true two-reservoir case. As a result, the effective temperature for the two electric fields  is $T^E_\text{eff}(t) = (e^2/\eta) \,(\la E_1^2\ra+ \la E_2^2\ra (1-\exp(-t/\tau))$, as shown in Fig.~\ref{fig2} (inset) for realistic trap parameters. Again, the decoherence properties can be understood with the help of the effective temperature: faster (slower) decoherence corresponds to higher (lower) effective temperature, as compared to the equal amplitude $\la E_1^2\ra=\la E_2^2\ra$ situation. The fact that $T^E_\text{eff}(t)$ increases in the latter indicates that the noisy fields heat the system up, a direct consequence of the non-Markovian nature of the environment ($T^E_\text{eff}$ is a constant in the limit $\tau \rightarrow 0$).  We also mention  that the exact \eqref{8} and approximate \eqref{10}  peak-to-peak ratios are indistinguishable for the parameters of Fig.~\ref{fig2}. 
 We can now readily generalize these results to arbitrary complex situations: Given an unknown number of noise sources, with unknown amplitudes and correlation times, we can use the  measurable effective temperature to predict the decoherence rate of the system via formula \eqref{10}. 

 An important property of the effective temperature is that it can be measured \cite{her02,abo04,aba08,ouk09}. We next discuss an experimental  scheme that allows   its determination  in  ion traps.
We consider a single ion coupled to a nonequilibrium amplitude reservoir  generated by an unspecified  number of  electric fields with different noise amplitudes and correlation times (noisy electrical electrode potentials  are the main sources of trap decoherence  \cite{des06}). The corresponding effective temperature \eqref{7} is  given by the ratio of the noise correlation function and the friction kernel. Both quantities can be determined experimentally by  measuring the motional heating rate of the trap \cite{tur00}. The latter is related to the power spectrum of the  fluctuation force, $S(\omega) = \int dt\, \exp(i\omega t) C(t)$, by $\dot n
\simeq e^2S(\omega)/(4m\hbar\omega)$, where $e$ is the charge of the ion \cite{tur00}. The noise correlation function can hence be directly obtained  from measurements of the heating rate for different  trap frequencies $\omega$. A first set of measurements of the heating rate for the  ambient equilibrium reservoir alone yields the  friction kernel $\eta$ (which in this case has been shown to be frequency independent \cite{tur00}) via the fluctuation-dissipation relation, $S(\omega) = 2D = 2 \eta T$.  A second set of measurements of the  heating rate in the presence of the random electric fields further provides the correlation function $C(t)$ of the nonequilibrium engineered reservoir and, in turn, the time-dependent effective temperature.

To conclude, the present findings show that nonequilibrium decoherence induced by an unknown number of different noise sources can be successfully described within a unified framework, in a way similar to equilibrium decoherence. They further highlight the central role of the effective temperature to characterize, understand and predict the loss of coherence of a quantum system in a general nonequilibrium environment.

 We thank G. Huber and F. Schmidt-Kaler for useful discussions. This work was  supported by the Emmy Noether Program of the DFG (LU1382/1-1) and the
cluster of excellence Nanosystems Initiative Munich (NIM).



\begin{references}

\bibitem{giu96} D. Giulini  {\it et al.}, {\it {D}ecoherence and the {A}ppearence of a {C}lassical {W}orld in {Q}uantum {T}heory}, (Springer, Berlin, 1996).

\bibitem{sch07} M.A. Schlosshauer, {\it {D}ecoherence and the {Q}uantum-{To}-{C}lassical {T}ransition}, (Springer, Berlin,  2007).

\bibitem{zur91} W. H. Zurek, Phys.
Today 44, 36Ð44 (1991).

\bibitem{zur03} W.~H. Zurek,  {Rev. Mod. Phys.} {\bf 75}, 715 (2003).

\bibitem{bru96} M. Brune {\it et al.}, Phys. Rev. Lett. {\bf 77}, 4887 (1996).

\bibitem{mya00} C.~J. Myatt {\it et al.}, Nature  {\bf 403}, 269 (2000).

\bibitem{nie00} M.A. Nielsen and I.L. Chuang. {\it {Q}uantum Computation and Quantum Information}, (Cambridge, Cambridge, 2000).

\bibitem{cug93} L.~F. Cugliandolo and J. Kurchan, Phys. Rev. Lett. {\bf 71}, 173 (1993).

\bibitem{cug97} L.~F. Cugliandolo, J. Kurchan and L. Peliti, Phys. Rev. E {\bf 55}, 3898 (1997).

\bibitem{kie02} D. Kielpinski, C. Monroe and D.~J. Wineland, Nature {\bf 417}, 709 (2002).

\bibitem{des06} L. Deslauriers {\it et al.}, Phys. Rev. Lett. {\bf 97}, 103007 (2006).

\bibitem{her02} D. H\'erisson and M. Ocio,
Phys. Rev. Lett. {\bf 88}, 257202 (2002).

\bibitem{abo04} B. Abou and F. Gallet,
Phys. Rev. Lett. {\bf 93}, 160603 (2004).

\bibitem{aba08} A.~R. Abate and D.~J. Durian, Phys. Rev. Lett. {\bf 101}, 245701 (2008).

\bibitem{ouk09} H. Oukris and N.~E. Israeloff, Nature Phys. {\bf 6}, 135 (2010).
\bibitem{poy96} J.~F. Poyatos, J.~I. Cirac and P. Zoller, Phys. Rev. Lett. {\bf 77}, 4728 (1996).

\bibitem{wei93} U. Weiss. {Q}uantum Dissipative Systems, (World Scientific, Singapore, 1993).



\bibitem{cal83} A.O. Caldeira and A.J.  Leggett, Ann. Phys. (N.Y.)  {\bf 149}, 374 (1983).

\bibitem{rei01}    P. Reimann, Chem. Phys. {\bf 268}, 337 (2001).

\bibitem{cal51} H.~B. Callen and J.~A. Welton, Phys. Rev. {\bf 83}, 34 (1951).

\bibitem{zam05} F. Zamponi {\it et al.}, 
 J. Stat. Mech.: Theory and Experiment, {\bf P09013}, (2005).

\bibitem{cug98} L. F. Cugliandolo and G. Lozano, Phys. Rev. Lett. {\bf 80}, 4979 (1998).


\bibitem{hub08} G. Huber {\it et al.}, New J. Phys. {\bf 10}, 013004 (2008).


\bibitem{paz93} J.~P. Paz, S. Habib and W.~H. Zurek, Phys. Rev. D {\bf 47}, 488 (1993).


\bibitem{lei03} D. Leibfried {\it et al.}. Rev. Mod. Phys. {\bf 75}, 281 (2003).

\bibitem{tur00} Q.~A. Turchette {\it et al.}, Phys. Rev. A {\bf 62}, 053807 (2000).





\end{references}
\end{document}